\documentclass[11pt]{article}

\usepackage{amsmath,dsfont}
\usepackage[english]{babel}
\usepackage{amsfonts}
\usepackage{amsmath,amssymb}
\usepackage{graphicx}

\makeatletter

\def\Z{\mathbb Z}
\def\R{\mathbb R}

\def\eqnarray{\stepcounter{equation}\let\@currentlabel=\theequation
\global\@eqnswtrue
\global\@eqcnt\z@\tabskip\@centering\let\\=\@eqncr
$$\halign to \displaywidth\bgroup\@eqnsel\hskip\@centering
  $\displaystyle\tabskip\z@{##}$&\global\@eqcnt\@ne
  \hfil$\displaystyle{\hbox{}##\hbox{}}$\hfil
  &\global\@eqcnt\tw@ $\displaystyle\tabskip\z@
  {##}$\hfil\tabskip\@centering&\llap{##}\tabskip\z@\cr}
\@addtoreset{equation}{section}
  \def\theequation{\thesection.\arabic{equation}}
\makeatother

\usepackage{amsmath,amssymb}
\def\beq{\begin{equation}}
\def\eeq{\end{equation}}
\def\beqa{\begin{eqnarray}}
\def\eeqa{\end{eqnarray}}
\def\barray{\begin{array}}
\def\earray{\end{array}}

\begin{document}

\title{
{\bf  The origin of the hidden supersymmetry }}

\author{\bf
{ V\'{\i}t Jakubsk\'y$^a$,  Luis-Miguel Nieto$^{b}$ and Mikhail S. Plyushchay${}^{b,c}$}\\
[4pt] {\small \textit{${}^{a}$ Nuclear Physics Institute,
ASCR,
250 68 \v Re\v z, Czech Republic}}\\
 {\small
\textit{${}^{b}$ Departamento de F\'{\i}sica Te\'orica,
At\'omica y \'Optica, Universidad de Valladolid,
47071, Valladolid, Spain}}\\
{\small \textit{${}^{c}$ Departamento de F\'{\i}sica,
Universidad de Santiago de Chile, Casilla 307, Santiago 2,
Chile  }}\\
 \sl{\small{E-mails:
 v.jakubsky@gmail.com, luismi@metodos.fam.cie.uva.es, mplyushc@lauca.usach.cl}}}
\date{}

\maketitle

\begin{abstract}
The hidden supersymmetry and related tri-supersymmetric
structure of the free particle system, the Dirac delta
potential problem and the Aharonov-Bohm effect (planar,
bound state, and tubule models) are explained by a special
nonlocal unitary transformation, which for the usual $N=2$
supercharges has a nature of Foldy-Wouthuysen
transformation. We show that in general case, the bosonized
supersymmetry of nonlocal, parity even systems emerges in
the same construction, and explain the origin of the
unusual $N=2$ supersymmetry of electron in
three-dimensional parity even magnetic field. The
observation extends to include the hidden superconformal
symmetry.
\end{abstract}

\vskip.5cm\noindent

\section{Introduction}

Some quantum systems possess a hidden symmetry associated
with  nontrivial integrals of motion, which reflect their
peculiar properties. A hidden supersymmetry \cite{SUSYbos}
was revealed recently in a class of quantum mechanical
systems with a local Hamiltonian. The list of such systems
includes the Dirac delta potential problem \cite{CP1}, the
Aharonov-Bohm effect (bound state \cite{CP1} and planar
\cite{AB1} models), the finite-gap periodic quantum
systems, and their infinite period limit in the form of
reflectionless systems \cite{CNP1,AdS}. All the listed
systems possess a degeneration in the energy spectrum
associated with a (twisted) parity symmetry. The hidden
supersymmetry of the first two systems is characterized by
the linear in the momentum supercharge operators; in the
last two families, the hidden supersymmetry is related to
the higher derivative nontrivial operator of the Lax pair
of the associated nonlinear integrable system. A usual
$N=2$ superextension of all these systems is accompanied by
a rich tri-supersymmetric structure rooted in the hidden
supersymmetry \cite{delta,finite,AB2}. \vskip0.08cm
\noindent \textit{A natural  question arises whether the
hidden and usual supersymmetry are somehow
related.}\vskip0.05cm

In this paper we show how the hidden supersymmetry and the
associated tri-supersymmetric structure  originate from the
usual $N=2$ supersymmetry and the (twisted) parity
symmetry. The observation is illustrated by the  models of
the Dirac delta potential problem and the Aharonov-Bohm
(AB) effect. We also discuss the nature of the earlier
revealed \emph{bosonized} supersymmetry of nonlocal
spinless quantum systems with parity even potentials
\cite{SUSYbos}, that appears in the same construction, and
explain the origin of the unusual $N=2$ supersymmetry of
electron in three-dimensional parity-even magnetic field
\cite{GK1,GK2}. Finally, we indicate that the observation
extends to include the hidden superconformal symmetry
\cite{AB1}, \cite{COP}.

\section{One-dimensional case: special unitary transformation}

Consider an $N=2$ supersymmetric one-dimensional quantum
mechanical system  \cite{Wit,GK2,SUSYQM}. It is described
by the Hamiltonian
\begin{equation}\label{H0}
    H=P^2 + W^2+\sigma_3W',
\end{equation}
and supercharges
\begin{equation}\label{Q120}
    Q_1=\sigma_1 P +\sigma_2 W,\qquad
    Q_2=i\sigma_3Q_1=-\sigma_2P+\sigma_1W,
\end{equation}
where $P=-i\frac{d}{dx}$, $W=W(x)$ is a superpotential,
$W'={dW}/{dx}$, $2m=1$ and $\hbar=1$.
 The $H$ and $Q_a$,
$a=1,2$, generate the $N=2$ supersymmetry,
\begin{equation}\label{susy0}
    \{Q_a,Q_b\}=2\delta_{ab}H\,,\qquad
    [Q_a,H]=0\,,
\end{equation}
for which the integral $\Gamma=\sigma_3$ plays a role of
the grading operator, $[\Gamma,H]=\{\Gamma,Q_a\}=0$.

Assume that the superpotential is an odd function,
$W(-x)=-W(x)$. Then the Hamiltonian is the even operator.
The reflection (parity) $\mathcal{R}$,
$\mathcal{R}x=-x\mathcal{R}$, $\mathcal{R}^2=1$, is the
additional, nonlocal integral of motion,
$[\mathcal{R},H]=0$. It anticommutes with the supercharges,
$\{\mathcal{R}, Q_a\}=0$. Let us realize a unitary
transformation,
\begin{equation}\label{U}
    \mathcal{O}\rightarrow
\tilde{\mathcal{O}}=U\mathcal{O}U^{-1}\,,\qquad
    U=\exp(i\pi S_-\Pi_-)=S_+ + \mathcal{R}S_-=\left(\begin{array}{cc}
                    \Pi_+&\Pi_-\\\Pi_-&\Pi_+\end{array}
\right),
\end{equation}
where $S_\pm=\frac{1}{2}(1\pm \sigma_1)$ and
$\Pi_{\pm}=\frac{1}{2}(1\pm \mathcal{R})$ are the
projectors\footnote{Transformation (\ref{U}) with
$\mathcal{R}$ changed for $-\mathcal{R}$ works as well, and
will be important for the planar AB effect.}. The
(nonlocal) operator (\ref{U}) satisfies
$U^\dagger=U^{-1}=U$, so that $U^2=1$. We have
    $\tilde{x}=x\sigma_1$,
    $\tilde{P}=P\sigma_1$,
$\tilde{\mathcal{R}}=\mathcal{R}$,
$\tilde{\sigma}_1=\sigma_1$,
    $\tilde{\sigma}_2=\sigma_2\mathcal{R}$,
    $\tilde{\sigma}_3=\sigma_3\mathcal{R}$, and
the transformed Hamiltonian and supercharges take a
\emph{diagonal} form,
\begin{equation}\label{H'}
    \tilde{H}=P^2+W^2+\sigma_3 \mathcal{R}W'\,,
\end{equation}
\begin{equation}\label{Q'}
    \tilde{Q}_1=P-i\sigma_3 \mathcal{R}W\,,\qquad
    \tilde{Q}_2=W+i\sigma_3 \mathcal{R}P\,.
\end{equation}
For the first order supercharge operators (\ref{Q120}),
this transformation has a nature of Foldy-Wouthuysen
transformation. We trade the locality of the operators for
their diagonal form. The transformed operators (\ref{H'})
and (\ref{Q'}) satisfy the same $N=2$ superalgebra,
\begin{equation}\label{susy0til}
    \{\tilde{Q}_a,\tilde{Q}_b\}=2\delta_{ab}\tilde{H}\,,\qquad
    [\tilde{Q}_a,\tilde{H}]=0\,,
\end{equation}
for which
\begin{equation}\label{G'}
    \tilde{\Gamma}=\sigma_3 \mathcal{R}
\end{equation}
plays a role of the grading operator.

Notice that the unitary transformation (\ref{U}) mediates
the intertwining relation
$U\mathcal{O}=\tilde{\mathcal{O}}U$ between the
corresponding Hamiltonians, supercharges and grading
operators.

In general, the transformed Hamiltonian (\ref{H'}) differs
from the original, \emph{local} Hamiltonian (\ref{H0}).
Though by the construction the both are unitary equivalent,
the Hamiltonian (\ref{H'}) is \emph{nonlocal}  due to the
presence of the reflection operator in the last term. There
are particular cases, however, for which the nonlocality is
suppressed by a specific choice of the superpotential, and
$\tilde{H}=H$. We will discuss some of such systems later
in the text.

The operators (\ref{Q120}) are not integrals of motion for
the transformed Hamiltonian (\ref{H'}), while the
transformed supercharges (\ref{Q'}) do not commute with the
initial Hamiltonian (\ref{H0}). At the same time, the three
operators $\sigma_3$, $\mathcal{R}$ and
$\mathcal{R}\sigma_3$ are the integrals for both $H$ and
$\tilde{H}$. The supercharges $Q_a$ commute with
$\sigma_3\mathcal{R}$, while the transformed supercharges
$\tilde{Q}_a$ commute with $\sigma_3$. Both the original
and the transformed supercharges anticommute with
$\mathcal{R}$. It is worth to note that there exists no
unitary transformation that would transform $\sigma_3$ (or
$\mathcal{R}\sigma_3$) into $\mathcal{R}$.

 The transformed system (\ref{H'}) can
be reduced to any of the two eigensubspaces of  $\sigma_3$.
Each of the obtained \emph{spinless} nonlocal systems,
\begin{equation}\label{H'eps}
    \tilde{H}_s=P^2+W^2+s \mathcal{R}W'\,,
    \quad {\rm where}\quad
    s=+1\quad {\rm or}\quad s=-1\,,
\end{equation}
still possesses a \emph{bosonized} $N=2$ supersymmetry
described by the \emph{nonlocal} supercharges,
\begin{equation}\label{Q'eps}
    \tilde{Q}_{1,s}=P-is \mathcal{R}W\,,\qquad
    \tilde{Q}_{2,s}=W+is \mathcal{R}P=is
    \mathcal{R}\tilde{Q}_{1,s}\,.
\end{equation}
The operator $\mathcal{R}$ plays the role of the grading
operator for both ($s=\pm 1$) reduced systems. Such
nonlocal supersymmetric systems were investigated in
\cite{SUSYbos}. Here, we just illustrate a general
situation by a simple example of the super-oscillator
system given by $W(x)=x$, see Fig. 1. In this case the
reduced  Hamiltonians can be presented in the form
$\tilde{H}_+=2(N+\Pi_+)$ ($s=+1$) and
$\tilde{H}_-=2(N+\Pi_-)$ ($s=-1$), where $N=a^+a^-$ is a
number operator, $\Pi_\pm=\frac{1}{2}(1\pm \mathcal{R})$
are the projectors on subspaces with even and odd
eigenvalues of $N$, and the reflection operator,
$\mathcal{R}=(-1)^N=\cos {\pi N}$, being written in the
coordinate representation with
$N=\frac{1}{2}(-\frac{d^2}{dx^2} +x^2-1)$, reveals a
\emph{nonlocal} nature of the supersymmetric systems
$\tilde{H}_+$ and $\tilde{H}_-$.

\begin{figure}[h!]\begin{center}
\includegraphics[scale=0.8]{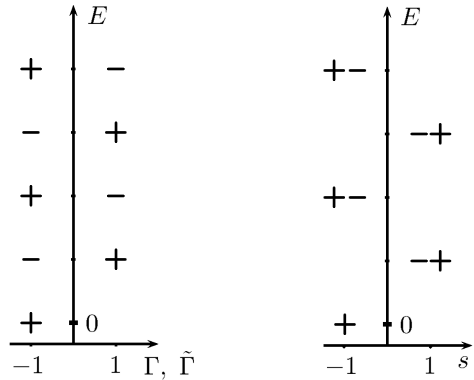}
\caption{The spectrum of the unitary equivalent
super-oscillator Hamiltonians $H$  and $\tilde{H}$ is shown
on the left; the eigenvalues of the grading operators
$\Gamma=\sigma_3$ and $\tilde{\Gamma}=\sigma_3\mathcal{R}$
are indicated below the corresponding states. The
``$\mathbf{+}$'' denotes the states with even parity, by
the ``$\mathbf{-}$'' we marked the states with odd parity.
On the right, the spectra of $\tilde{H}_{s}$, $s=\pm 1$,
are shown. The degeneracy of the energy levels in each
subsystem, reflected by the hidden supersymmetry, is
manifested. The hidden supersymmetry is exact for $s=-1$
(there is a singlet ground state) whereas it is broken for
$s=1$.    }
\end{center}
\end{figure}

\section{Special one-dimensional cases}

Consider now the special cases when the transformed
Hamiltonian coincides with the original one. This happens
when (\ref{U}) is the additional integral, $[H,U]=0$, of
the $N=2$ supersymmetric system.

\subsection{Free particle on a line}

We start with the simplest case which corresponds to a free
particle, $W(x)=0$, $H=P^2$, as it sheds a light on general
features of the supersymmetric structure associated with
the hidden supersymmetry in the systems we consider in what
follows.

For the free particle, both pairs of the operators,
(\ref{Q120}) and (\ref{Q'}), are integrals of motion. For
$\Gamma=\sigma_3$ chosen as the grading operator, the
$Q_1=\sigma_1P$ and $Q_2=-\sigma_2P$ are the odd, fermionic
integrals, while the $\tilde{Q}_1=P$ and
$\tilde{Q}_2=i\sigma_3\mathcal{R}P$ are the even, bosonic
integrals. Accordingly, the relations (\ref{susy0}) have to
be supplied with the commutation relations
\begin{equation}\label{QQtil}
    [Q_a,\tilde{Q}_b]=0\,,\qquad
    [\tilde{Q}_1,\tilde{Q}_2]=-2i\mathcal{R}\sigma_3 H\,.
\end{equation}

The $\Gamma=\mathcal{R}\sigma_3$ can be identified as the
grading operator as well. The  integrals (\ref{Q'}) play
then the role of the fermionic supercharges which satisfy
the relations (\ref{susy0til}) (with $\tilde{H}=H$), while
(\ref{Q120}) are the bosonic integrals. Relations
(\ref{QQtil}) are changed for the relations of a similar
form with the duality-like replacement $Q_a\leftrightarrow
\tilde{Q}_a$, $\mathcal{R}\sigma_3\leftrightarrow
\sigma_3$.

If the parity operator is identified as the grading
operator, $\Gamma=\mathcal{R}$, all the integrals $Q_a$ and
$\tilde{Q}_a$ should be treated as fermionic supercharges.
Then the anticommutation relations (\ref{susy0}) and
(\ref{susy0til}) are supplemented with the relations
\begin{equation}\label{QtilQH}
    \{Q_a,\tilde{Q}_b\}=2(-1)^a\left(-\delta_{b1}
    \sigma_a+\delta_{b2}\epsilon_{ac}\sigma_c\mathcal{R}
    \right)H\,,
\end{equation}
which just reflect the fact that the sigma matrices are the
even integrals of motion that have to be treated as the
even generators,
$[\sigma_a,\sigma_b]=2i\epsilon_{ab}\sigma_3$, in the
complete {nonlinear tri-supersymmetry}, see \cite{delta}.

The reduction of the $N=2$ supersymmetric structure
generated by $H$ and $\tilde{Q}_a$ to the eigensubspaces
$\sigma_3=+1$ and $\sigma_3=-1$ results in the bosonized
supersymmetry, in which  the $\Gamma=\mathcal{R}$ is
identified as the grading operator, and $\tilde{Q}_1=P$ and
$\tilde{Q}_2=i\mathcal{R}P$ (the sign in definition of the
latter operator  is irrelevant) play the role of the
fermionic supercharges. The eigenstates of the
$\tilde{Q}_2$ are
$\psi_k(x)=\frac{1}{\sqrt{2}}e^{-i\pi/4}(e^{ikx}+ie^{-ikx})=\cos
kx +\sin kx$, $\tilde{Q}_2\psi_k(x)=k\psi_k(x)$, $k\in
(-\infty,\infty)$, cf. the eigenstates $e^{ikx}$ of
$\tilde{Q}_1$, $\tilde{Q}_1e^{ikx}=ke^{ikx}$. Notice also
that
 $\tilde{Q}_\pm=\tilde{Q}_1\pm i\tilde{Q}_2=P\Pi_\pm$
 realize the Darboux transformation between the eigenstates
 $\cos kx$ and $\sin kx$, $k\geq 0$,
 of the free particle Hamiltonian $H=P^2$\,: $\tilde{Q}_+\cos
 kx=-k\sin kx$, $\tilde{Q}_+\sin
 kx= 0$, $\tilde{Q}_-\cos
 kx=0$, $\tilde{Q}_-\sin
 kx=k\cos kx$.

\subsection{ Dirac delta potential problem}

Besides a free particle case in $\R^1$ with $W(x)=0$, let
us mention another simple but nontrivial model on the line,
for which the unitary transformation $U$ is the symmetry of
Hamiltonian. It is given by
\begin{equation}\label{Wdel}
    W(x)=\beta \varepsilon(x),
\end{equation}
where $\varepsilon(x)$ is a sign function defined as
$\varepsilon(x)=1$ for $x\geq 0$, and $\varepsilon(x)=-1$
for $x<0$. In this case Hamiltonian (\ref{H0}) corresponds
to the $N=2$  superextended Dirac delta potential problem
\cite{Boya},
\begin{equation}\label{Hdel}
    H=P^2+\beta^2+2\beta\sigma_3\delta(x)\,.
\end{equation}
Since
$\mathcal{R}\delta(x)\psi(x)=\delta(-x)\psi(-x)=\delta(x)\psi(x)$,
the transformed Hamiltonian (\ref{H'}) coincides with the
original one, (\ref{Hdel}). Similarly to the free particle
case, $[Q_a,\tilde{Q}_b]=0$.

After reduction to the eigensubspaces of the diagonal
integral $\sigma_3$, we get two spinless one-dimensional
Dirac delta potential problems with the hidden $N=2$
supersymmetry, described by
\begin{equation}\label{HredDel}
    \tilde{H}_s=P^2+\beta^2+s\, 2\beta\delta(x)\,,\qquad
    s=+1\quad {\rm or} -1\,,
\end{equation}
\begin{equation}\label{QredDel}
    \tilde{Q}_{1,s}=P+s\, i\beta\varepsilon(x)\mathcal{R}\,,\qquad
    \tilde{Q}_{2,s}=i
    \mathcal{R}\tilde{Q}_{1,s}\,.
\end{equation}
The hidden supersymmetry of the spinless systems
(\ref{HredDel}) and the tri-supersymmetric structure of the
spin-$1/2$ system (\ref{Hdel}) were studied in \cite{CP1},
\cite{delta}. Here we just notice that while the
Hamiltonian (\ref{HredDel}) is local, the both supercharges
(\ref{QredDel}) of the hidden supersymmetry are non-local
operators. For $\beta>0$ and $s=-1$ (the case of the
attractive delta function potential), the system has a
singlet bound state of zero energy separated by the energy
gap $\beta^2$ from the doubly degenerate continuous
(scattering) part of the spectrum, i.e. corresponding
hidden supersymmetry is unbroken. For $\beta>0$, $s=+1$
(repulsive delta function potential), the system is
characterized by the broken $N=2$ bosonized supersymmetry
that reflects coherently the double degeneration of all the
(scattering) states with $E>\beta^2$ in the spectrum of the
system.

\subsection{Bound state Aharonov-Bohm model}

Consider a charged \emph{spinless} particle subjected to
move on a unit circle $x^2+y^2=1$ (placed in the plane
$z=0$) in the presence of the magnetic field of a flux
line,
$B_z(x,y)=\epsilon_{ij}\partial_{i}A_j=\Phi\delta(x,y)$.
The Hamiltonian of the system is given by
\begin{equation}\label{HAB}
    H_\alpha=(p_\varphi+\alpha)^2\,,
\end{equation}
where $p_\varphi=-i\frac{d}{d\varphi}$, $\varphi$ is the
angular variable on a unit circle, and
$\alpha=-\frac{e}{2\pi c}\Phi$. This configuration
corresponds to the bound state Aharonov-Bohm effect
\cite{ABbound}.

The usual $N=2$ supersymmetric extension is similar to that
of the free one-dimensional particle discussed above, with
the change $P\rightarrow p_\varphi +\alpha$. The analogue
of the parity integral $\mathcal{R}$, however, does not
exist for arbitrary values of the rescaled magnetic flux
parameter $\alpha$.

Consider  a twisted reflection operator
\begin{equation}\label{Rtwist}
 \mathcal{R}=e^{-2i\alpha\varphi}R_\varphi\,,
\end{equation}
where the $R_\varphi$ is a reflection in $\varphi$,
$R_\varphi\psi(\varphi)=\psi(-\varphi)$. Operator
(\ref{Rtwist}) is well defined (maps $2\pi$-periodic
functions into $2\pi$-periodic ones), and commutes with the
Hamiltonian (\ref{HAB}) \emph{only when} $\alpha$ takes
\emph{integer} or \emph{half-integer} values.

The discrete spectrum of the system (\ref{HAB}) with the
energy levels $E_l=(l+\alpha)^2$, $l=0,\pm 1,\pm 2,\ldots$,
which correspond to the states $e^{il\varphi}$, has a
degeneration typical for the $N=2$ supersymmetry only in
the same cases $\alpha=n$, or $\alpha=n+\frac{1}{2}$,
$n\in\Z$. For $\alpha=n$, the system is unitary equivalent
to the free particle on a circle case  ($\alpha=0$) since
$p_\varphi+n=\mathcal{U}_np_\varphi\mathcal{U}_n^{-1}$,
$\mathcal{U}_n=e^{-in\varphi}$. The zero-energy ground
state ($l=-n$) is nondegenerate while the states with
$l=k\neq -n$ and $l=-k-2n$ form a doublet of the same
energy (not taking into account a double degeneration of
all the levels related to the decoupled spin variables). On
the contrary, for $\alpha=n+\frac{1}{2}$, all the energy
levels are positive and doubly degenerate modulo the
degeneration associated with the spin degrees of freedom :
$E_l=E_{-l-2n-1}=(l+n+\frac{1}{2})^2\geq 1/4$.

Hence, the procedure of the special unitary transformation
and  subsequent reduction applies in the current system as
well, where it relates the earlier observed hidden
supersymmetry of the bound state AB effect \cite{CP1} with
the usual $N=2$ supersymmetry associated with the decoupled
spin degrees of freedom.

\section{Generalization to the two dimensions}

Consider a charged spin-1/2 particle confined in the plane
in the presence of the perpendicular magnetic field, that
is described by the Pauli Hamiltonian
\begin{equation}\label{H2d}
    H=\mathcal{P}_i^2 -\frac{e}{c}\sigma_3 B,
\end{equation}
where
$\mathcal{P}_i=-i\partial_i-\frac{e}{c}A_i(\mathbf{x})$,
$i=1,2$, and $B(\mathbf{x})=\epsilon_{ij}\partial_i
A_j(\mathbf{x})$. For arbitrary magnetic field, such a
system possess the $N=2$ supersymmetry (\ref{susy0}) [with
$\Gamma=\sigma_3$] generated  by the supercharges
\cite{AhCas}
\begin{equation}\label{Q2d}
    Q_1=\sigma_i \mathcal{P}_i\,,\qquad
    Q_2=\epsilon_{ij}\sigma_i \mathcal{P}_j=i\sigma_3Q_1\,.
\end{equation}
As we shall see, the application of this simple but
\emph{formal} construction of the  $N=2$ supersymmetry is
accompanied by the proper definition of the involved
operators in the case of the planar AB effect \cite{AB2}.

Assume now that the magnetic field is an even function,
$B(-\mathbf{x})=B(\mathbf{x})$, described in terms of the
odd vector potential, $A_i(-\mathbf{x})=-A_i(\mathbf{x})$.
Then the system (\ref{H2d}) will have an additional,
nonlocal integral
\begin{equation}\label{RLpi}
    \mathcal{R}=\exp(i\pi L)\,,
\end{equation}
$[H,\mathcal{R}]=0$, which corresponds to a rotation in
$\pi$, where $L=-i\epsilon_{ij}x_i\partial_j$ is the
orbital angular momentum. The operator (\ref{RLpi})
satisfies the relations $\mathcal{R}x_i=-x_i\mathcal{R}$,
$\mathcal{R}^2=1$, and, therefore, supercharges (\ref{Q2d})
are the parity-odd operators, $\{\mathcal{R},Q_a\}=0,$
$a=1,2$. Then we can apply the analysis of Section 2 based
on the special unitary transformation, in which the
operator $\mathcal{R}$ is given by (\ref{RLpi}). The
supercharges (\ref{Q2d}) can be obtained alternatively by
making the changes $P\rightarrow \mathcal{P}_1$,
$W\rightarrow \mathcal{P}_2$ in (\ref{Q120}). The
transformed supercharges $\tilde{Q}_1$ and $\tilde{Q}_2$
take the form
\begin{equation}\label{QPi}
    \tilde{Q}_1=\mathcal{P}_1-i\sigma_3\mathcal{R}
    \mathcal{P}_2\,,\qquad
    \tilde{Q}_2=\mathcal{P}_2+i\sigma_3\mathcal{R}
    \mathcal{P}_1\,.
\end{equation}

Likewise in the one-dimensional systems, the operator $U$
of the unitary transformation does not commute with the
Hamiltonian in general. However, there are exceptional
cases, including the case of the free particle ($A_i=B=0$).
Let us comment on this case briefly here. Following the
discussion of Section 3.1, we get an explanation for the
hidden $N=2$ supersymmetry of the free spinless planar
particle system\,: it can be related to the $N=2$
 supersymmetry of the spin-1/2 analog of the system via the
special unitary transformation (\ref{U}) and subsequent
reduction to any of the two eigensubspaces $\sigma_3=+1$ or
$\sigma_3=-1$. In the free particle case, the generators of
the hidden supersymmetry,
\begin{equation}\label{Qfreevec}
    \tilde{Q}_i=\mathcal{P}_i-i\sigma_3\mathcal{R}
    \epsilon_{ij}\mathcal{P}_j,
\end{equation}
form a two-dimensional vector with respect to to the total
angular momentum $\mathcal{J}=L+\frac{1}{2}\sigma_3$,
$[\mathcal{J},\tilde{Q}_i]=i\epsilon_{ij}\tilde{Q}_j$, in
contrast with the scalar supercharges $Q_a$,
$[\mathcal{J},Q_a]=0$.

Below, we shall ellaborate another two-dimensional systems
where the hidden supersymmetry can be related to the
standard $N=2$ supersymmetry via the unitary
transformation. At first, we will analyze the
two-dimensional system which is a symbiosis of the bound
state AB model considered in Section 3.3, and of the free
particle -- the particle on the cylinder. The second model
will be the celebrated planar AB model.

\subsection{Aharonov-Bohm effect\,: the tubule model}

Consider the model of a charged \emph{spin-1/2} particle on
the cylinder in the presence of the AB flux along the
symmetry axis ($x_1=0$, $-\infty<x_2=y<\infty$) of the
cylinder. It is described by
\begin{equation}\label{HABtub}
    H=(p_\varphi +\alpha)^2 +p_y^2\,.
\end{equation}
The (singular) magnetic field is not orthogonal to the
two-dimensional surface here, but (\ref{HABtub}) is
obtained from (\ref{H2d}) by changing
$\mathcal{P}_1\rightarrow p_\varphi+\alpha$,
$p_\varphi=-i\partial/\partial\varphi$, and
$\mathcal{P}_2\rightarrow p_y=-i\partial/\partial y$, and
by omitting the spin term there.
 The supercharge integrals are obtained then from
(\ref{Q2d}) by the same change,
\begin{equation}\label{QABtub}
     Q_1=\sigma_1(p_\varphi+\alpha)+\sigma_2 p_y\,,
     \qquad Q_2=i\sigma_3Q_1\,.
\end{equation}
 As in
the case of the bound-state AB model, for integer and
half-integer values of the rescaled magnetic flux $\alpha$,
the Hamiltonian (\ref{HABtub}) has an additional integral
\begin{equation}\label{Rtub}
    \mathcal{R}=e^{-2i\alpha\varphi}R_\varphi R_y\,,\qquad
    \alpha=n\quad {\rm or}\quad n+\frac{1}{2}\,,
\end{equation}
where $R_y$ is the operator of reflection in the $y$
coordinate, $R_y y=-yR_y$. The integral (\ref{Rtub})
anticommutes with both supercharges $Q_a$. The additional,
commuting with $Q_a$ integrals,
\begin{equation}\label{QtilABtub}
    \tilde{Q}_1=(p_\varphi+\alpha)-i
    \sigma_3\mathcal{R}p_y,\qquad
    \tilde{Q}_2=i\mathcal{R}\sigma_3\tilde{Q}_1,
\end{equation}
are obtained from (\ref{QPi}) via the indicated above
substitution, i.e. by applying  the unitary transformation
(\ref{U}) to (\ref{QABtub}). The tri-supersymmetric
structure associated with the three possible choices for
the grading operator can  be computed following the line of
Section 3.1.

\subsection{Planar Aharonov-Bohm effect}

Consider the $N=2$ supersymmetric system that corresponds
to the planar AB effect \cite{AB0} for the spin$-1/2$
particle. This system is described by the Hamiltonian
(\ref{H2d}) with the electromagnetic potential given by
\begin{equation}\label{3}
    \vec{A}=\frac{\Phi}{2 \pi}
    \left(-\frac{x_2}{x_1^2+x_2^2},
    \frac{x_1}{x_1^2+x_2^2}\right)=
    \frac{\Phi}{2 \pi r}\left(-\sin\varphi \, ,\
    \cos \varphi \right)\,,
\end{equation}
where we use the polar coordinates, $x_1=r\cos\varphi, \
x_2=r\sin\varphi$. Potential (\ref{3}) corresponds to the
singular magnetic field,
 $B(\mathbf{x})=\Phi\,\delta^2(x_1,x_2)$. The explicit form of the
Hamiltonian is
\begin{equation}\label{H}
  \mathcal{H}_\alpha=
            -\partial^2_r -\frac{1}{r}\partial_r+\frac{1}{r^2}(-i
            \partial_\varphi+\alpha)^2 +\alpha
            \frac{1}{r}\delta(r)\sigma_3\,,\quad
            \alpha=-\frac{e}{2\pi c}\Phi\,,
\end{equation}
where we use the identity $\delta^2(x_1,x_2)=\frac{1}{\pi
r}\delta(r)$ for the two dimensional Dirac delta function.
Since the vector potential and magnetic field are singular
functions at the point $\mathbf{x}=0$, the appropriate
domains have to be specified for the Hamiltonian and
supercharges (\ref{Q2d}) in order to keep them well defined
(self-adjoint).

The AB system with the integer value of the magnetic flux
is unitary equivalent to the free-particle case
($\alpha=0$) which was discussed above. In general, the
relation $\mathcal{H}_{\alpha+n}=U_n\mathcal{H}_\alpha
U_n^{-1}$ with  $U_n=e^{-in\varphi}\mathds{1}$, where
$\mathds{1}$ is the unit $2\times 2$ matrix, tells that we
can assume $\alpha\in (0,1)$ without loss of generality. As
it was shown in \cite{AB2}, the supercharges of the $N=2$
supersymmetry are well defined in two cases only, which
correspond to two different self-adjoint extensions of the
Hamiltonian $\mathcal{H}_{\alpha}$, denoted as
$\mathcal{H}^{0}_{\alpha}$ and
$\mathcal{H}^{\pi}_{\alpha}$, cf. (\ref{Rtub}). In other
words, there are just two self-adjoint extensions of the
Hamiltonian that are consistent with the $N=2$
supersymmetry.
 These two self-adjoint extensions,
\begin{equation}\label{H0pi}
     \mathcal{H}^{\gamma=0}_{\alpha}=
     \left(\begin{array}{cc}H_{\alpha}^0&0\\
     0&H_{\alpha}^{AB}\end{array}\right)\,,\qquad
     \mathcal{H}^{\gamma=\pi}_{\alpha}=
     \left(\begin{array}{cc}H_{\alpha}^{AB}&0\\
     0&H_{\alpha}^{\pi}\end{array}\right),
\end{equation}
differ in their domains. They are well defined on the
locally square integrable functions, that are regular at
the origin up to a single partial wave, where the singular
behavior is enforced. The two component wave functions from
the domain of $\mathcal{H}^{\gamma}_{\alpha}$ have to
comply with the following boundary conditions\,:
\begin{equation}\label{boundary}
     \lim_{r\rightarrow 0^+}\Psi\sim\left(\begin{array}{l}
    (1+e^{i\gamma})2^{-\alpha}\Gamma(1 -
    \alpha) r^{-1+\alpha}e^{-i\varphi}\\
    (1-e^{i\gamma})2^{-1+\alpha}\Gamma(\alpha)
     r^{-\alpha}
    \end{array}
    \right).
\end{equation}
Explicit form of the corresponding supercharges defined on
the same domain is
\begin{equation}\label{Q12sic}
     Q_1^\gamma=
     \left(\begin{array}{cc}
     0&{\cal P}_1-i{\cal P}_2\\
     {\cal P}_1+i{\cal P}_2&0\end{array}\right)\,,\qquad
     Q_2^\gamma=i\sigma_3Q_1^\gamma=
     \left(\begin{array}{cc}
     0&{\cal P}_2+i{\cal P}_1\\
     {\cal P}_2-i{\cal P}_1&0
     \end{array}\right)\,.
\end{equation}
Note that as formal differential operators, the
supercharges are the same for both values of $\gamma$;
however, for $\gamma=0$ and $\gamma=\pi$, their domains are
different. The same is valid for the operators
$H^0_\alpha$, $H^\pi_\alpha$ and $H^{AB}_\alpha$. For the
first two, the corresponding domains admit singular (at
zero) wave functions in corresponding partial waves, while
the domain of $H^{AB}_\alpha$ includes only regular at zero
functions. Therefore, the two Hamiltonian operators
(\ref{H0pi}) describe the two different systems.

The both systems (\ref{H0pi}) have additional, nonlocal
integral of motion (\ref{RLpi}) which, unlike the bound
state AB effect and the related tubule model, exists for
\emph{arbitrary} value of the flux parameter [remind that
we restrict $\alpha\in(0,\pi)$], and acts here on the
angular variable as
$\mathcal{R}\varphi\mathcal{R}=\varphi+\pi$.

Define the two different unitary operators,
\begin{equation}\label{Upm}
    U_{\pm}=\left(\begin{array}{cc}
                    \Pi_{\pm}&\Pi_{\mp}\\\Pi_{\mp}&\Pi_{\pm}\end{array}
\right),
\end{equation}
which satisfy the relations $U_\pm^\dagger=U_\pm$,
$U_+^2=U_-^2=1$, and
$
    U_+\sigma_3U_+=\sigma_3\mathcal{R}$,
$    U_-\sigma_3U_-=-\sigma_3\mathcal{R}$. Operator $U_+$
corresponds here to  (\ref{U}), while $U_-$ is obtained
from it via the change $\mathcal{R}\rightarrow
-\mathcal{R}$. Both $U_{\pm}$ commute with the formal
Hamiltonian operator (\ref{H}). It is necessary, however,
to check how they act on the wave functions from the domain
of $\mathcal{H}^{\gamma}_{\alpha}$. The $U_+$ respects the
boundary conditions (\ref{boundary}) if and only if
$\gamma=\pi$, while $U_-$ does not alter (\ref{boundary})
for $\gamma=0$. The domain of $\mathcal{H}^{\pi}_{\alpha}$
($\mathcal{H}^{0}_{\alpha}$)
  is invariant with respect to $U_+$ ($U_-$), and
  therefore
$
 [U_+,\mathcal{H}^{\pi}_{\alpha}]=0$,
 $[U_-,\mathcal{H}^{0}_{\alpha}]=0$.
Under the unitary transformation $U_+$ ($U_-$), the
Hamiltonian $\mathcal{H}^{\pi}_{\alpha}$
($\mathcal{H}^{0}_{\alpha}$) remains the same.

Unitary transformation of the supercharges ${Q}_{a}^{\pi}$
(${Q}_{a}^{0}$) by the $U_+$ ($U_-$) gives  the
corresponding supercharges of the hidden $N=2$
supersymmetry. They can be written in the unified form
\begin{equation}\label{hiddenSUSY}
     \tilde{Q}^{\gamma}_1=\left(\begin{array}{cc}
    \mathcal{P}_1+ie^{i\gamma}\mathcal{R}\mathcal{P}_2&0\\0&\mathcal{P}_1-
    ie^{i\gamma}R\mathcal{P}_2
     \end{array}\right)\,,
     \qquad \tilde{Q}^{\gamma}_2
    =-ie^{i\gamma}\sigma_3\mathcal{R}\tilde{Q}_1^{\gamma}\,,\qquad
    \gamma=0,\pi\,.
\end{equation}
Like in the free planar particle case, the supercharges
$Q_a^\gamma$ of the usual $N=2$ supersymmetry are scalars
with respect to the total angular momentum ${\cal
J}=L+\frac{1}{2}\sigma_3$, while the generators of the
hidden supersymmetry, $\tilde{Q}^{\gamma}_1$ and
$\tilde{Q}^{\gamma}_2$, for both $\gamma$ values form a two
dimensional vector,
    $[{\cal
    J},\tilde{Q}^\gamma_i]=i\epsilon_{ij}\tilde{Q}_j^\gamma$.
This also follows from the alternative representation of
(\ref{hiddenSUSY}),
\begin{equation}\label{Qtildvec}
    \tilde{Q}^\gamma_i={\cal P}_i\cdot\mathds{1}
    +ie^{i\gamma}\mathcal{R}\sigma_3\cdot\epsilon_{ij}{\cal
    P}_j\,,
\end{equation}
cf. (\ref{QPi}).  Reduction to the eigensubspaces
$\sigma_3=+1$ and $\sigma_3=-1$ produces the three
different  AB models for a scalar particle described by the
Hamiltonians $H^0_\alpha$, $H^\pi_\alpha$ and
$H^{AB}_\alpha$, each of which possesses the hidden
supersymmetry generated by the corresponding diagonal
component of (\ref{Qtildvec}). This explains the origin of
the hidden supersymmetry in the AB effect for the scalar
particle that was observed  in \cite{AB1}. Notice that the
generators of the usual supersymmetry, $Q_a^\gamma$,
commute with the generators $\tilde{Q}^\gamma_i$ of the
hidden supersymmetry,
$[{Q}_{a}^{\gamma},\tilde{Q}^\gamma_i]=0$, for both
$\gamma=0$ and $\gamma=\pi$.

The tri-supersymmetry of the system, associated with three
alternative grading operators and discussed in~\cite{AB2},
can be obtained in the same vain as in Section 3.1.

\section{Unusual $N=2$ supersymmetry in the three dimensions}

Consider a three-dimensional spin-$1/2$ particle in
magnetic field
$B_i(\mathbf{x})=\epsilon_{ijk}\partial_jA_k(\mathbf{x})$.
The system is described by the  Hamiltonian,
\begin{equation}\label{H3D}
    H=P_i^2+\sigma_iB_i\,,
\end{equation}
and possesses the $N=1$ supersymmetry described by the
supercharge
\begin{equation}\label{Q3D}
    \tilde{Q}_1=P_i\sigma_i\,,
\end{equation}
$\tilde{Q}_1^2=H$. Here $P_i=-i\partial_i
-\frac{e}{c}A_i(\mathbf{x})$, and summation in $i=1,2,3$ is
assumed.

The $N=1$ supersymmetry can be extended to the
\emph{artificial} $N=2$ supersymmetry by introducing the
``isospin'' degrees of freedom described by another set of
Pauli matrices, which we denote by  $\Sigma_l$, $l=1,2,3$,
and by defining
\begin{equation}\label{QQS}
    Q_1=\Sigma_1\tilde{Q}_1,\qquad
    Q_2=\Sigma_2\tilde{Q}_1=i\Sigma_3 Q_1\,.
\end{equation}

Suppose now that the vector potential
$\mathbf{A}(\mathbf{x})$ is a parity odd function,
$A_i(-\mathbf{x})=-A_i(\mathbf{x})$. Then magnetic field is
an even function, the parity operator $\mathcal{R}$,
$\mathcal{R}x_i=-x_i\mathcal{R}$, anticommutes with the
supercharges $Q_a$, and commutes with $H$. The structure we
have obtained is similar to the supersymmetric structure of
the one-dimensional free particle with the $\tilde{Q}_1$
and $\Sigma_l$ corresponding here to the $P$ and $\sigma_i$
in the latter system.

Realizing the unitary transformation (\ref{U}) (with
$\sigma_1$ substituted for $\Sigma_1$), and subsequently
reducing the system to the eigensubspace $\Sigma_3=+1$, we
find that the system (\ref{H3D}) is described by the $N=2$
supersymmetry with the supercharges (\ref{Q3D}) and
$\tilde{Q}_2=i\mathcal{R}\tilde{Q}_1$, for which the parity
$\mathcal{R}$ plays a role of the grading operator. This
shows that the unusual $N=2$ supersymmetry of the system
(\ref{H3D}) with odd vector potential, observed earlier in
\cite{GK1,GK2}, has the same nature as the hidden
supersymmetry of the free particle.

\section{Discussion}

Up to now, our discussion was restricted to the
supersymmetries generated by the time-independent
operators. In the case of the spin-1/2 free particle and
the planar Aharonov-Bohm model, the $N=2$ supersymmetry can
be extended to the superconformal symmetry, supplying the
Hamiltonian with bosonic generators of the dilatations $D$
and special conformal transformations $K$. Their commutator
with the the supercharges $Q_a$ generate the additional odd
integrals $S_a$, that depend explicitly on time
\cite{AB2,PAH}, $S_a=i[K,Q_a]$. Since the indicated bosonic
generators $K$ and $D$ are diagonal operators and commute
with the reflection operator $\mathcal{R}$, they are
invariant with respect to the unitary transformation $U$.
This is not the case for $S_a$, which is transformed into
the diagonal time-dependent symmetry
$\tilde{S}_a=U\,S_a\,U^{-1}$. The subsequent reduction to
the eigensubspaces $\sigma_3=+1$ and $\sigma_3=-1$ gives
rise to the hidden superconformal symmetry of the scalar
free particle \cite{COP} and for the spinless AB effect
\cite{AB1} and, therefore, clarifies its origin.

In all the systems we considered, the generators of the
usual $N=2$ supersymmetry commute with the generators of
the hidden supersymmetry. This means that if one of the
generators of the usual supersymmetry is identified as a
first order Hamiltonian like that in the massless Dirac
particle case \cite{Jack,CDP}, such a first order system
will possess a hidden $N=2$ supersymmetry. This observation
can be applied in the condensed matter systems described by
the Dirac-Weyl equation, and will be elaborated elsewhere.

We have explained the origin of the hidden supersymmetry of
some quantum mechanical systems, where the corresponding
supercharges are the first order (nonlocal) differential
operators. Notice that this construction, based on  the
nonlocal unitary Foldy-Wouthuysen transformation, is
completely different from that in \cite{AS}, where the
hidden supersymmetry is described by local supercharges.
The open question is then whether a usual $N=2$ linear or
nonlinear supersymmetry of the quantum periodic finite-gap
systems \cite{Fin1,Fin2,KhS,Fin3,Fin4} could be related in
a similar way, via a nonlocal unitary transformation, to
the hidden supersymmetry associated with the higher order
nontrivial Lax operators \cite{finite}. \vskip0.3cm

\noindent \textbf{Acknowledgements.}
 The work of MSP has been partially supported by
 FONDECYT Grant 1095027, Chile and  by Spanish Ministerio de
 Educaci\'on under Project
SAB2009-0181 (sabbatical grant). LMN has been partially
supported by the Spanish Ministerio de Ciencia e
Innovaci\'on (Project MTM2009-10751) and Junta de Castilla
y Le\'on (Excellence Project GR224). VJ was supported by the
Czech Ministry of Education, Youth and Sports within the
project LC06002.


\end{document}